\documentclass[aps,twocolumn,superscriptaddress]{revtex4}
\usepackage{graphicx}
\usepackage{epsfig}

\begin{document}
\title{A unifying approach to left handed material design}

\author{Jiangfeng Zhou}
\affiliation{Department of Electrical and Computer Engineering
and Microelectronics Research Center,Iowa State University, Ames, Iowa 50011}%

\author{Eleftherios N. Economon}
\affiliation{Institute of Electronic Structure and Laser - FORTH,and
Department of Physics, University of Crete, Greece}

\author{Thomas Koschny}
\author{Costas M. Soukoulis}
\affiliation{Ames Laboratory and Department of Physics and
Astronomy,Iowa State University, Ames, Iowa 50011}
\affiliation{Institute of Electronic Structure and Laser - FORTH,and
Department of Materials Science and Technology, University of Crete,
Greece}

\begin{abstract}
In this letter we show that equivalent circuits offer a qualitative
and even quantitative simple explanation for the behavior of various
types of left-handed (or negative index) meta-materials. This allows
us to optimize design features and parameters, while avoiding trial
and error simulations or fabrications. In particular we apply this
unifying circuit approach in accounting for the features and in
optimizing the structure employing parallel metallic bars on the two
sides of a dielectric film.
Pacs: {42.70.Qs, 41.20.Jb, 42.25.Bs, 73.20.Mf}
\end{abstract}


\maketitle

\begin{figure}[htb]\centering
  \includegraphics[width=8cm]{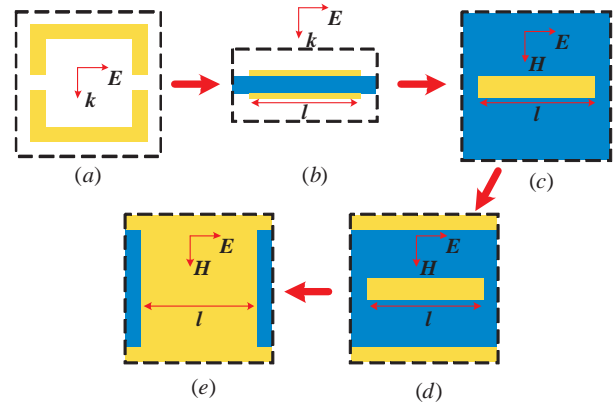}\\
  \caption{(Color online) The two cut single metallic SRR (a) can be transformed to a pair of parallel metallic
  bars separated by a dielectric (b, view in ($\vec{E},\vec{k}$) plane; c, view in ($\vec{E},\vec{H}$) plane).
  By adding continuous wires,
  design d (view in ($\vec{E},\vec{H}$) plane) results, which can be modified to a fully
  connected one on both sides of the thin dielectric board (e). The
  dashed square defines the unit cells with dimension $a_x$ (parallel to
  $\vec{H}$), $a_y$ (parallel to $\vec{E}$) and $a_z$ (parallel to
  $\vec{k}$).
  \label{fig1_srr2cpw}}
\end{figure}

Left-handed materials exhibit a negative permeability, $\mu$, and
permittivity, $\epsilon$, over a common frequency range
\cite{Veselago_1968}. Negative permeability is the result of a
strong resonance response to an external magnetic field; negative
permittivity can appear either by a plasmonic or a resonance
response (or both) to an external electric field. Negative $\mu$ and
negative $\epsilon$ lead to negative index of refraction, $n$, and
to a left-handed triad of $\vec{k}, \vec{E}, \vec{H}$; hence, the
names negative index materials (NIMs) or Left-handed Materials
(LHMs). Pendry\cite{PRL_pendry_wire_plasma,IEEE_pendry_srr}
suggested a double metallic split-ring resonator (SRR) design for
negative $\mu$ and a parallel metallic wire periodic structure for
an adjustable plasmonic response. Several variation of the initial
design have been studied; among them a single ring resonator with
several cuts has been proved capable of reaching negative $\mu$ at
higher frequency \cite{PRL_zhou_freq_sat}; in Fig.
\ref{fig1_srr2cpw}(a) a two cut single ring is shown schematically.
This, by a continuous transformation, can be reduced to a pair of
carefully aligned metal bars separated by a dielectric spacer of
thickness $t_s$\cite{PRB_zhou_cwp,optleter_shalaev_2005}; in Figs
\ref{fig1_srr2cpw}(b) and \ref{fig1_srr2cpw}(c) the view in the
($\vec{E},\vec{k}$) and ($\vec{E},\vec{H}$) planes of this structure
is shown together with the directions of $\vec{k}, \vec{E}, \vec{H}$
of the incoming $EM$ field.

\begin{figure}[htb]\centering%

\includegraphics[width=8cm]{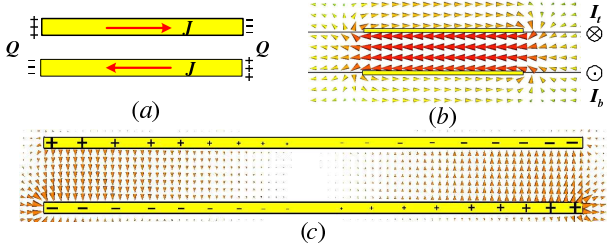}\\%
\caption{(Color online) At the magnetic resonance the currents (a,
in ($\vec{E},\vec{k}$) plane, view in $\vec{H}$ direction), the
magnetic field (b, in ($\vec{H},\vec{k}$) plane) and the electric
field (c, in ($\vec{E},\vec{k}$) plane) are shown. Sizes of the
cones show the intensity of magnetic field $\vec{H}$ (b) and
electric field $\vec{E}$ (c) in logarithm scale. \label{fig_field}}
\end{figure}

The design shown in Figs \ref{fig1_srr2cpw}(b,c), besides its
simplicity, has distinct advantages over conventional SRRs. The
incident electromagnetic wave is normal to the structure as shown in
Fig. \ref{fig1_srr2cpw}(b), which enable us to build NIMs by only
one layer of sample and achieve relatively strong response.
Conventional SRRs, although they exhibit magnetic resonance which
may produce negative $\mu$, they fail to give negative $\epsilon$ at
the same frequency range and, hence, they are incapable by
themselves to produce NIMs. An extra continuous wire is needed to
obtain negative $\epsilon$ via plasmonic response
\cite{PRL_pendry_wire_plasma,PRL_Smith_first_NIM}. In contrast, the
pair of parallel metallic plates is expected to exhibit not only a
magnetic resonance [Fig. \ref{fig_field}(c), antisymmetric mode],
but to show an electric resonance as well [symmetric mode] properly
located in frequency by adjusting the length, $l$, of the pair.

The simulations were done with the CST Microwave Studio (Computer
Simulation Technology GmbH, Darmstadt, Germany) using the lossy
metal model for copper with a conductivity $\sigma=5.8\times10^7$
for a single unit cell with periodic boundary in the (E,H) plane,
field distribution and scattering amplitudes have been calculated.
The $\epsilon, \mu$ in Fig. \ref{fig_retrieval} have been obtained
by a retrieval procedure \cite{PRB_smith_retrieval_2002}. At the
magnetic resonance the two plates sustain anti-parallel currents
producing a magnetic field $\vec{B}$ confined mainly in the space
between the plates and directed opposite to that shown in Fig.
\ref{fig1_srr2cpw}(c); the electric field, because of the opposite
charges accumulated at the ends of the two plates, is expected to be
confined within the space between the plates and near the end
points. Indeed, detailed simulations, shown in Fig.
\ref{fig_field}(c), confirm this picture.

%

At the electric resonance the currents at the two bars are parallel
(symmetric mode); the magnetic field lines go around both bars,
while the electric field is mostly confined in the space between the
nearest neighbor edges of the two pairs of bars belonging to
consecutive unit cells.

%
\begin{figure}[htb]\centering
  \includegraphics[width=7cm]{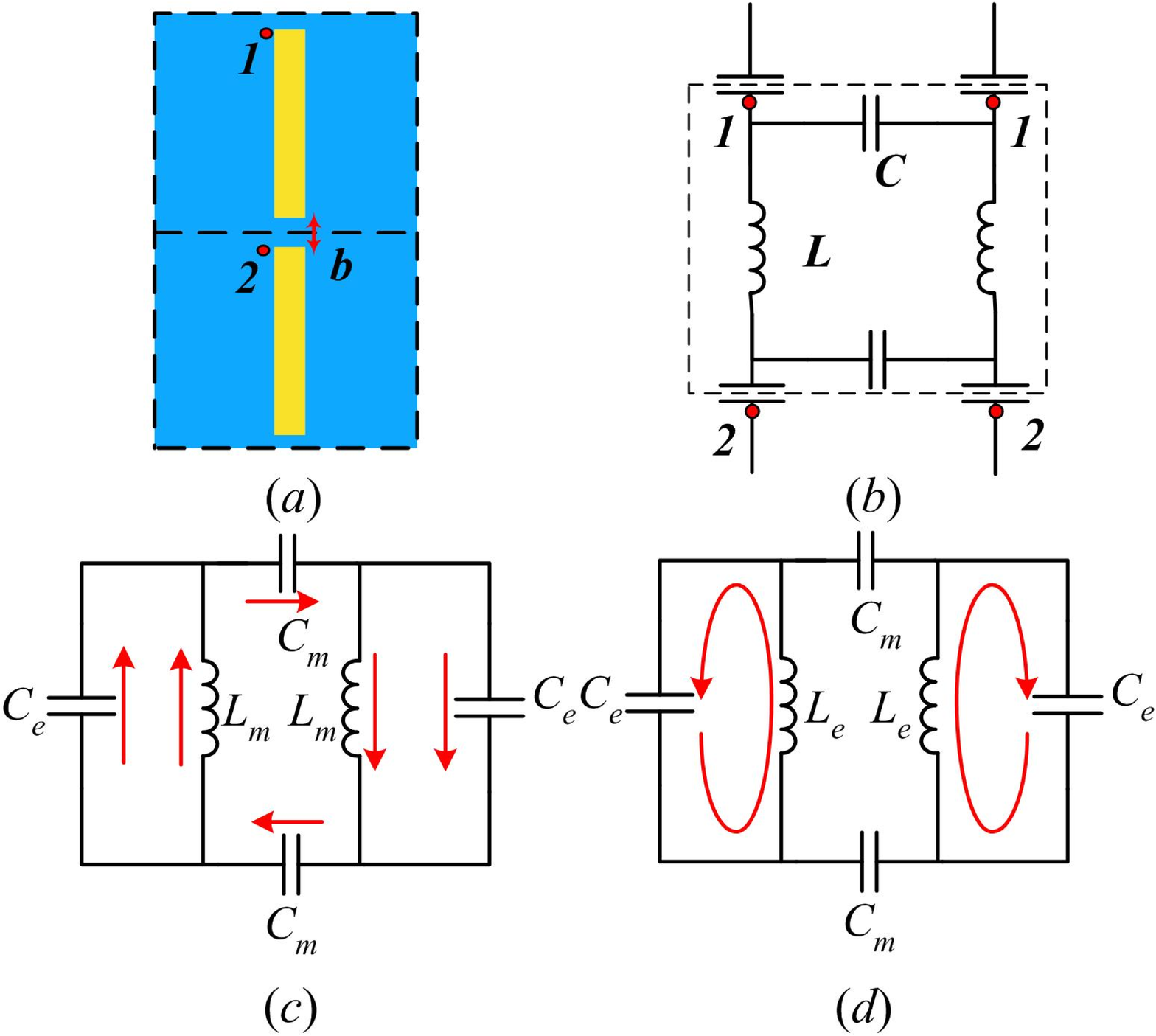}\\
  \caption{(Color online) Current distribution of the two parallel
  metallic bar design (a) (side view, the parallel plates are behind one another) can be accounted for by the equivalent
  circuit (b), which, since points 1 and 2 are equivalent because of
  the periodicity, reduces to circuit (c) and (d) for the magnetic
  (c) and electric (d) resonance respectively.
  \label{fig_circuit}}
\end{figure}

The field and current configurations for both the antisymmetric and
the symmetric mode can be accounted for by equivalent $R,C,L$
circuits as shown in Fig. \ref{fig_circuit} (in which for simplicity
the resistor elements have been omitted). Near the magnetic
resonance frequency where the current configuration is as in Fig.
\ref{fig_circuit}(c), the magnetic field is between the two plates
and it is, to a good approximation, uniform (Fig.
\ref{fig_field}(b)). Hence the total inductance $L$, as calculated
by the magnetic field energy, is
\begin{equation}\label{eqn_L_plate}
L=2L_m\simeq\mu\frac{t_{s}}{w}l,
\end{equation}
where $l$ is the length of the wire, $t_s$ is the thickness of the
dielectric spacer and $w$ is the width of the wire.

 Notice that at telecommunication or optical frequencies, where the linear dimension
are in the tens or hundreds of nm, the kinetic energy of the
drifting electrons makes a contribution comparable or larger than
the magnetic energy. Hence, another additional inductance must be
added to the right hand side of Eq.\ref{eqn_L_plate}
\cite{PRL_zhou_freq_sat}.

Each of the capacitance $C_m$ must be given by a formula of the type
\begin{equation}\label{eqn_C_m}
C_m=\frac{\epsilon wl^{'}}{t_s},
\end{equation}
where by inspection of Fig. \ref{fig_field}(c), $l^{'}=c_1l$ with
the numerical factor $c_1$ in the range $0.2\leq c_1 \leq 0.3$.  The
capacitance $C_e$ can be approximated by that of two parallel wires
of radius $t_m$ and length $w$ at a distance $b$ apart
\begin{equation}\label{eqn_C_e}
C_e=\frac{\pi\epsilon w}{\ln(b/t_m)},
\end{equation}
where $t_m$ is the thickness of each metallic bar and $b$ is the
separation of neighboring pairs Fig. \ref{fig_circuit}(a,b)
($b=a_y-l$). The magnetic resonance frequency, $\omega _m$, is
obtained by equating the impedance $Z$ (of $L_m$ and $C_e$ in
parallel) with minus the impendence $-i/C_m\omega$ of the
capacitance $C_m$.

\begin{figure}[htb]\centering
  \includegraphics[width=5cm]{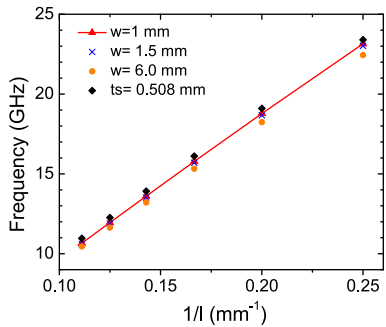}\\
  \caption{(Color online) Linear dependence of the magnetic resonance frequency, $f_m$, as obtained by
  simulation, on the inverse length $l$; this result as well as its
  independence on $w$ and $t_s$ is in agreement with the simple
  formula (\ref{eqn_f_m}). ($t_s=0.254mm$ for triangular, cross, circle;
  $w=1mm$ for diamond; and for all cases, $b=0.5 \sim 5.5mm$, $a_x=20mm$).
  \label{fig_dep_l}}
\end{figure}

Since $Z=iL_m\omega/(1-L_mC_e\omega^2)$ we obtain
\begin{equation}\label{eqn_omega_m}
\omega_m=\frac{1}{\sqrt{L_m(C_m+C_e)}}\simeq\frac{1}{\sqrt{L_mC_m}}.
\end{equation}
The last relation follows because, for the values we have used
($l=7\mathrm{mm}, w=1\mathrm{mm}, t_s=0.254\mathrm{mm},
t_m=10\mathrm{\mu m}$ and $b=0.3\mathrm{mm}$), $C_e\simeq 0.1C_m$.
Combining the Eq.\ref{eqn_L_plate} and Eq.\ref{eqn_C_m} we find that

\begin{equation}\label{eqn_f_m}
f_m=\frac{\omega_m}{2\pi}=\frac{1}{2\pi l\sqrt{\epsilon\mu}\sqrt{c_1/2}}%
=\frac{1}{2\pi\sqrt{c_1\epsilon_r/2}}\frac{c}{l}
\end{equation}
where $\epsilon_r=2.53$ is the reduced dielectric constant of the
dielectric, $\epsilon_r=\epsilon/\epsilon_0$. In Fig.\ref{fig_dep_l}
we compare our result of Eq.\ref{eqn_f_m}, which shows that $f_m$ is
a linear function of only $1/l$, with detailed simulations results.
Fig. \ref{fig_dep_l} shows the dependence of the magnetic resonance
frequency as obtained from the retrieved resonant effective $\mu$ on
the inverse length of the parallel metallic bars (Fig.
\ref{fig1_srr2cpw}b), for different widths $w$ (separation between
parallel bars $t_s=0.254mm$ fixed) and two different separations
$t_s$ (width of bars $w=1mm$ fixed). Complete quantitative agreement
is obtained if $c_1=0.22$. Notice the independence of the simulation
results on the width $w$ and the dielectric thickness $t_s$. It is
worthwhile to point out that the result $f_m\sim1/l$ is robust over
a wide range of parameters even if Eqs. \ref{eqn_L_plate} and
\ref{eqn_C_m} are not valid. To see this point, consider the extreme
case of a pair of thin wires (as opposed to a pair of bars) of
length $l$, cross-section radius $r$, at a distance $d$ apart such
that $r\ll d\ll l$. For such a system
$L=(\mu/4\pi)[1+4\ln(d/r)]\simeq(\mu l/\pi)\ln(d/r)$ and
$C\simeq\epsilon\pi l/\ln(d/r)$. Thus again $f_m\sim
1/\sqrt{\epsilon\mu}l$.
%

For frequencies near the electric resonance, because of mirror
symmetry in Fig. \ref{fig_circuit}(d), there is no current passing
through the capacitances $C_m$. As a results the electric resonance
frequency $f_e$ is given by $f_e=1/(2\pi\sqrt{C_e L_e})$,
%
%
where $L_e$ is expected to be of the form $(\mu /\pi)g(w/l)$ where
$g(x)$ is a function which for $x\rightarrow0$ behaves as $-\ln(x)$.

\begin{figure}[htb]\centering
  \includegraphics[width=5cm]{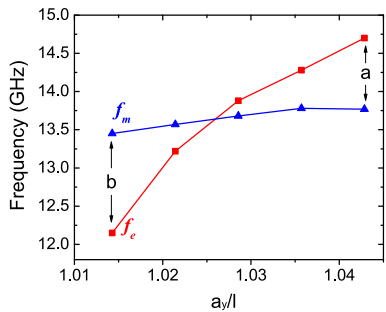}\\
  \caption{(Color online) Magnetic resonant frequency $f_{m}$ cross over with electrical %
  resonant frequency $f_{e}$ as $a_y/l=1+b/l$ varies between $7.1\mathrm{mm}$
  and $7.3\mathrm{mm}$; $a_x=20\mathrm{mm}$.
  \label{fig_crossing}}
\end{figure}

\begin{figure}[htb]\centering
  \includegraphics[width=7cm]{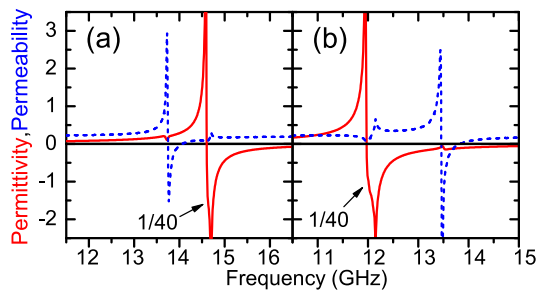}\\
  \caption{(Color online) Retrieved $\epsilon_{\mathrm{eff}}$ (solid lines) and $\mu_{\mathrm{eff}}$ (dotted
  lines) for two cut wires.
  (a) and (b) correspond to points a ($a_y=7.3\mathrm{mm}$, $a_x=20\mathrm{mm}$) and b ($a_y=7.1\mathrm{mm}$,
  $a_x=20\mathrm{mm}$) in
  Fig.~\ref{fig_crossing}. Notice that both the response are Lorentz
  like.
  \label{fig_retrieval}}
\end{figure}

We point out that $f_e$ is a rather sensitive function of the small
distance $b$, because $C_e$ depends on $b$, while $L_e$ is
practically independent on $b$. Indeed the ratios $f_e(2b)/f_e(b)$
and $f_e(3b)/f_e(b)$ for $b=0.1\mathrm{mm}$ according to the
equation of $C_e$ and equation of $f_e$ are respectively $1.14$ and
$1.215$ in good agreement with the simulation results in Fig.
\ref{fig_crossing} ($1.13$ and $1.21$ respectively); the dependence
of both $f_m$ and $f_e$ on $a_y/l=1+b/l$ is shown in Fig.
\ref{fig_crossing}.

Fig. \ref{fig_crossing} in combination with Fig. \ref{fig_retrieval}
suggest the optimum design parameters for making the two bar scheme
to produce negative index $n$: One has to avoid the crossing region
where essentially to a considerably degree the two resonances cancel
each other. Since the electric resonance  is much stronger and,
hence, much wider we have to bring the magnetic resonance within the
negative region of $\epsilon$, i.e. we must have $f_e$ lower than
$f_m$ as in Fig. \ref{fig_retrieval}(b), rather that the other way
around, i.e. we must have
\begin{equation}\label{eqn_f_em}
\left(\frac{f_e}{f_m}\right)^2=\frac{L_m}{L_e}\left(1+\frac{C_m}{C_e}\right)<1
\end{equation}
This can be achieved by increasing $C_e$ either by decreasing $b$ or
by increasing at the ends of each bar the width $w$ choosing a
double $T$ shape for each bar \cite{APL_zhou_cwp}.

Still another possibility to make the negative $\epsilon$ region
wider (and more negative) is to add continuous metallic wires as in
Fig. \ref{fig1_srr2cpw}(d) which produce a plasmonic response
\cite{PRB_zhou_cwp}. By adjusting the width of these wires their
effective plasma frequency $f_p$ can be made larger than the
frequency, $f_1$, at which the continuous curve in Fig.
\ref{fig_retrieval}(b) crosses the axis ($f_1\simeq 16 GHz$).

Finally the width of the bars, $w$, can increase until the bars join
the "infinite" wires producing thus a continuous connected network
which can be constructed by opening periodically placed rectangular
holes on uniform metallic films covering both sides of a dielectric
sheet
\cite{science_soukoulis_2006,PRL_zhang_2005_037402,PRL_zhang_2005_137404}.

In this letter we have shown that $L,C$ equivalent circuits can
account for the $EM$ properties of various negative index artificial
meta-materials (NIMs), even at a quantitative level; furthermore,
this simple unifying circuit approach offers a clear guidance in
adjusting the design and optimizing the parameters for existing and
possibly, future NIMs.

We gratefully acknowledge the support of Ames Laboratory (operated
by Iowa State University under Contract No. W-7405-Eng-82), the
AFOSR under MURI grant (FA9550-06-1-0337), EU Network of Excellence
projects METAMORPHOSE and PHOREMOST, and Defence Advanced Research
Projects Agency (DARPA) contract HR0011-05-C-0068).



\bibliography{C:/zjf/research/paper/references/references}
\bibliographystyle{apsrev}

\end{document}